\documentclass[sigconf]{aamas} 

\usepackage{balance} 
\usepackage{xcolor}
\usepackage{nicefrac}
\usepackage{pdfpages}

\usepackage{times}
\usepackage{soul}
\usepackage{url}
\usepackage[utf8]{inputenc}
\usepackage{amsmath}
\usepackage{amsthm}
\usepackage{booktabs}
\usepackage{placeins}
\usepackage[ruled]{algorithm2e} 

\usepackage{xcolor}
\usepackage{nicefrac}
\usepackage{pdfpages}
\usepackage{multirow}
\usepackage{wrapfig}

\newcount\Comments  
\newcommand{\kibitz}[2]{\ifnum\Comments=1{\color{#1}{#2}}\fi}
\newcommand{\rmr}[1]{\kibitz{red}{[
Reshef says: #1]}}

\newcommand{\nm}[1]{\kibitz{orange}{[Nick says:#1]}}
\def\cite{\citep}

\newcommand{\labeq}[2]{\begin{equation}\label{eq:#1}#2\end{equation}}

\setcopyright{ifaamas}
\acmConference[AAMAS '23]{Proc.\@ of the 22nd International Conference
on Autonomous Agents and Multiagent Systems (AAMAS 2023)}{}
{London, UK}{? (eds.)}
\copyrightyear{2023}
\acmYear{2023}
\acmDOI{}
\acmPrice{}
\acmISBN{}

\acmSubmissionID{35}

\title[Mitigating Skewed Bidding for Conference Paper Assignment]{Mitigating Skewed Bidding for Conference Paper Assignment}

\author{Inbal Rozencweig}
 \affiliation{
  \institution{Technion - Israel Institute of Technology}
  \city{Haifa}
    \country{Israel}}
 \email{inbalroz91@gmail.com}

 \author{Reshef Meir}
 \affiliation{
  \institution{Technion - Israel Institute of Technology}
  \city{Haifa}
    \country{Israel}}
 \email{reshefm@ie.technion.ac.il}

 \author{Nicholas Mattei}
 \affiliation{
  \institution{Tulane University}
  \city{New Orleans, LA}
    \country{USA}}
 \email{nsmattei@tulane.edu}

 \author{Ofra Amir}
 \affiliation{
  \institution{Technion - Israel Institute of Technology}
  \city{Haifa}
    \country{Israel}}
 \email{oamir@technion.ac.il}

\begin{abstract}
The explosion of conference paper submissions in AI and related fields, has underscored the need to improve many aspects of the peer review process, especially the matching of papers and reviewers. Recent work argues that the key to improve this matching is to modify aspects of the \emph{bidding phase} itself, to ensure that the  set of bids over papers is balanced, and in particular to avoid \emph{orphan papers}, i.e., those papers that receive no bids. In an attempt to understand and mitigate this problem, we have developed a flexible bidding platform to test adaptations to the bidding process. Using this platform, we performed a field experiment during the bidding phase of a medium-size international workshop that compared two bidding methods. We further examined via controlled experiments on Amazon Mechanical Turk various factors that affect bidding, in particular the order in which papers are presented \cite{cabanac2013capitalizing,fiez2020super}; and information on paper demand \cite{meir2021market}. Our results suggest that several simple adaptations, that can be added to any existing platform, may significantly reduce the skew in bids, thereby improving the allocation for both reviewers and conference organizers.
\end{abstract}

\begin{document}

\maketitle



         


\section{Introduction}
Academic peer review of papers and grants sits at the heart of academic work and is the cornerstone of modern scientific enterprise \cite{Boha13a}. In some areas of computer science (mainly AI/ML), where most papers are submitted to large conferences, the fate of a paper is very much in the hands of automated assignment algorithms that help program chairs distribute thousands of papers among a similar number of committee members that serve as reviewers \cite{meir2021market}. For this matching to happen, the committee members must first submit their preferences over papers. These preferences are supposed to reflect both the competence and the interest of the reviewer in reviewing those particular papers, using a designated platform---a process typically referred to as \emph{bidding}, and that many of the readers probably know well from their own experience. \citet{cabanac2013capitalizing} provide a detailed account of conference bidding and review flow. After the bidding process, one of the many algorithms for matching under preferences \cite{manlove2013algorithmics,DBLP:reference/choice/KlausMR16} can be used to find an assignment satisfying various notions of optimality, fairness, stability, etc. \cite{chen2020stable,aziz2015fair,AzizHMHCRR,benabbou2021finding}.

Crucially, the current design of the bidding process falls far short of eliciting the full preferences and capabilities of reviewers. First, in some widely used platforms (e.g. EasyChair) there are only three levels of preference:  `no' / `maybe' / 'yes'. Other platforms provide a finer scale for reviewers to express their preferences. However, it is not clear to what extent reviewers use this flexibility, as \emph{extreme responding} or \emph{scale end bias} is a well known phenomena in many social sciences \cite{furnham1986response}. 
Additionally, it is not clear yet whether or not a finer grained scale of responses would actually lead to more desirable matchings between reviewers and papers.\rmr{is there literature on more and less expressive scales for preferences? e.g. showing that allowing a finer scale does not necessarily lead to collecting more information / introduces biases due to cultural/personal differences} \nm{I added a cite and am actually working with Nihar on some stuff in this space but there isn't anything out there at the moment.}
Second and more importantly, going over the entire list of submissions to determine the fit of every paper would take hours, whereas most reviewers would not invest that much time in bidding. Given that  modern computer science conferences may have thousands of papers submitted to them, automated systems are being increasingly used to impute the bids of reviewers over papers, an example being the Toronto Paper Matching System (TPMS) \cite{charlin2013toronto}.

Hence, for these reasons and many others, it has been claimed that \emph{skewed bidding}, i.e., where a few papers get many bids and some papers get no bids, is one of the main reasons for poor paper assignment \cite{fiez2020super,meir2021market,shah2021systemic,shahthewebconf,LianMNW18,leyton2022matching}. The argument is that some papers get insufficient (or no) bids  and have to be assigned randomly or manually by the program chair, often ending up at unqualified reviewers. For example, \citet{cabanac2013capitalizing} analyzed data from nearly 20,000 reviews in dozens of conferences managed on ConfMaster, and showed that more than 8,000 (42\%) were done by reviewers who did not bid on the paper at all!\footnote{Indeed, ConfMaster also allows reviewers to express negative preference on a paper by bidding `no', but this is not very helpful when facing thousands of papers.} A poor assignment, in turn, may affect  review quality \cite{stelmakh2019peerreview4all,peng2017time,RodriguezBS07}; and increase the overhead on conference chairs, who need to handle these \emph{orphan papers} that receive no bids via manual (re)assignments. Skewed bidding is also likely to put obstacles in the way of achieving alternative goals such as fairness~\cite{payan2021will,LianMNW18}, as creating a fair assignment crucially depends on actually knowing the preferences of the reviewers.

\paragraph{Skewed Bidding at AI Conferences}
 At AAMAS, where we have data from PrefLib \cite{MaWa13a,MaWa17}, there are also a high number of orphan papers.\footnote{Note that the AAMAS data reviewers were able to opt out of being included in the public dataset, hence some papers and bids are missing from this dataset.} The AAMAS 2015 dataset contains 9,817 bids of 201 reviewers over 613 papers; this represents about 40\% of the actual 22,360 bids of 281 reviewers over 670 papers. The 2016 data contains 161 out of 393 reviewers with bids over 442 out of 550 papers. Within this, for AAMAS 2015 papers had 6.9 bids on average, yet there are 30 papers that have no bids at all (5\%) and 95 papers that have less than 3 bids (15.4\%), while for AAMAS 2016 papers had 6.5 bids on average, but there are 8 papers that have no bids at all (1.8\%) and 54 papers with less than 3 bids (12.2\%).
 
Simply increasing the bidding requirement, which increases the burden on reviewers during the bidding process, may still not be sufficient to deal with the issue of orphan papers. For example, at IJCAI 2018 each paper received almost 40 bids \emph{on average} (!), and yet 140 papers (4\%) had only two or fewer bids~\cite{meir2021market}.

\subsection{Proposed Solutions}

Given the strong skew in bidding, there have been two recent suggestions to alleviate the problem of skewed bidding put forward in the literature:
\begin{enumerate}
 \item Presenting low-demand papers higher on the list~\cite{cabanac2013capitalizing,fiez2020super};
    \item Providing  information regarding paper demand~\cite{meir2021market}.
\end{enumerate}
Interestingly, the first suggestion builds on reviewers' cognitive biases, while the latter exploits their (bounded) rational behavior.

In more detail, \citet{fiez2020super} proposed an algorithm to determine the order in which papers are presented to the reviewer during bidding, taking advantage of the ordering of papers to bidders. This suggestion  rests on the \emph{primacy effect}: items that appear earlier on a list are more likely to be selected~\cite{murphy2006primacy}. Primacy effects have been empirically shown to occur in conference bidding data on ConfMaster~\cite{cabanac2013capitalizing}. The underlying idea is that demand can be smoothed by taking advantage of well known cognitive biases rather than providing more information to bidders.

The other suggestion, by \citet{meir2021market}, considers a model where the demand over papers is known (or revealed) to the bidders. They showed that as long as reviewers are individually rational and interpret their probability of being assigned a paper as inversely proportional to demand, a simple market-based scheme induces an incentive to follow the recommended instructions, and thereby reduces the skew in bids and leads to an improved assignment. Drawing inspiration from the Trading Post Mechanism~\cite{ShapleyShubik77}, they suggest tagging papers with their \emph{inverse price} rather than actual demand, and assign a \emph{budget} the bidder is encouraged to use. Interestingly, rational bidders then have an incentive to exhaust their budget, but some bias in favor of high-price (low-demand) papers is necessary to obtain more balanced bids. Thus the model predicts bounded rationality would lead to the best results.


In both the work of \citet{meir2021market} and \citet{fiez2020super}, the actual behavior of the individual bidder (i.e. how their bid is affected by order  or  demand) is \emph{assumed}, and the theoretical and empirical results are contingent on these assumptions. However, bidding behavior with prices has never been tried or empirically validated, and while primacy effect has been shown to exist on average, it is not well understood how substantial it is compared to other factors.

\subsection{Contribution}  
The goal of this paper is to explore how different components of the bidding platform affect the probability that a participant will select a particular paper. The main motivation, following~\cite{fiez2020super,meir2021market} is to promote the selection of papers with few bids, thereby reducing the skew and indirectly improving the paper assignment.

Since previous work has suggested to control either the order of papers~\cite{cabanac2013capitalizing,fiez2020super}, or the information given to users on the demand~\cite{meir2021market}, these are the main parameters we considered. 

\begin{description}
\item[Hypothesis~1 (Order Effect)] Subjects tend to select papers appearing earlier on the list. 
\item[Hypothesis~2 (Demand Effect)] Subjects tend to select papers that are indicated as low-demand.
\end{description}
In addition we are interested in how these tendencies, if they exist, are distributed in the population, as well as in various factors affecting them. Hence, we designed and executed two types of experiments. The first is a field experiment on a medium-size workshop, and the second is a large scale experiment on Amazon Mechanical Turk where we control all the variables. In both experiments only some of the subjects were exposed to information on the demand, so their behavior can be compared to the control group. 

Our main findings support both hypotheses, as we show that both paper order and information on demand can be used to shift reviewers towards low-demand papers. 
However at the individual level there is a substantial difference. The order of  papers has an effects on most subjects, but in a rather weak  manner. In contrast,  we identify in both experiments a small group of people that are \emph{highly sensitive} to the  demand,  and results from the field experiment suggest that their effect on the bid distribution is substantial. 
We further study via controlled experiments the relative and cumulative effect of exposing the subjects to different forms of information on the demand, and simple factors affecting compliance with the bidding instructions.
\rmr{how is that?}
We conclude with a list of simple, practical suggestions to improve the use bidding platforms so as to reduce the prevalent skew in paper bidding, thereby improving paper matching.

\subsection{Related Work}
\rmr{Other related work on paper assignment. Work from consumer research on primacy effect, on highlighting.}

Ordering effects are well studied in economic and psychological models of choice. Typically, decision makers attend to the first few and last few items in a list more than the rest, increasing response rates for these items \cite{krosnick1987evaluation}. In an academic context, papers appearing earlier on an email digest are more likely to be downloaded and cited~\cite{feenberg2017s}. \citet{cabanac2013capitalizing} were the first to show that ordering effects occur in paper bidding. Later, \citet{fiez2020super} suggested a sophisticated sorting algorithm that takes into account both dynamic demand and estimated reviewers' preferences.

\citet{RodriguezBS07} aimed at uncovering the factors underlying bidders' behavior in the JCDL'05 conference. Their starting point was that bids are expected to reflect the (objective) expertise of the reviewer w.r.t. the domain of the submission. They evaluate this expertise through alternative means, e.g., co-author network or keyword occurrence. The authors find very low correlation between reviewers' areas and their bids, and conjecture that \emph{reviewer fatigue} may be responsible. Our work does not get into whether reviewers' preferences are indeed based on expertise (as opposed to, say, curiosity and  interest in the title). It does however shed light on the other, more consistent factors that affect bidding behavior.

A major challenge in behavioral studies is having subjects with real-world preferences \emph{and} comparing behavior against true preferences, which are private. Ideally, we would combine these in a single experiment that cleverly elicits the real preferences, as in the work of \citet{budish2017can} on course allocation, or by performing individual  exit polls~\cite{blais2000calculus} on voters. Since there is no conference, let alone a large one, that uses a similar mechanism for paper bidding, we resorted to use a combination of field and controlled experiments.

\paragraph{Assignment Algorithms} The assignment of papers to reviewers is formally a version of the multi-agent resource allocation problem with capacities \cite{BoChLa16} and has been well studied in a number of areas of computer science \cite{GoSl07a,LianMNW18}, economics \cite{BuCa12a}, and beyond \cite{DPS14a}. \citet{garg2010assigning} provide a comprehensive discussion of assignment algorithms, their application to the review process, and different methods for evaluating the quality of an assignment from both the conference and reviewer standpoint. Two popular ways to evaluate assignments are maximizing either the egalitarian welfare \cite{DeHi88a}, i.e., making sure the worst off reviewers is as happy as can be or the utilitarian welfare, i.e., maximizing the sum of reported utilities for assigned papers across all reviewers. There are other refinements of these solution concepts \cite{garg2010assigning,LianMNW18}, and a large literature on calibrating feedback across reviewers for better assignment \cite{wang2019your}. While the workshop in which we ran our field experiment used the  utilitarian maximal assignment (an assignment maximizing social welfare), the results we report are independent of the assignment algorithm in use. Note that while assignment of heterogeneous tasks is also common in other domains such as crowdsourcing~\cite{assadi2015online}, the `workers' in paper bidding have some unique features. They are volunteers (which is also true in some crowdsourcing tasks), they often participate repeatedly every year, and they expect a roughly fixed workload.

Some modern platforms use TPMS or other systems that infer the interests of the reviewer from her list of publications or other sources~\cite{charlin2013toronto}. However it does not seem that implicit preferences induced from TPMS are less skewed than explicit bids. As \citet{fiez2020super} find in their study, TPMS scores result in a very skewed and sub-optimal bid distribution, where many papers receive very low scores. For example, in the TPMS dataset from ICLR 2018, out of the 911 papers, 85 of them (9.3\%) have a \emph{maximum} similarity score $\leq$ 0.1 (on a $[0,1]$ scale), meaning that these papers are very unlikely to get bids from reviewers.

\section{Experimental Design}\label{sec:design}
\begin{figure*}
    \centering
    \includegraphics[width=0.48\textwidth]{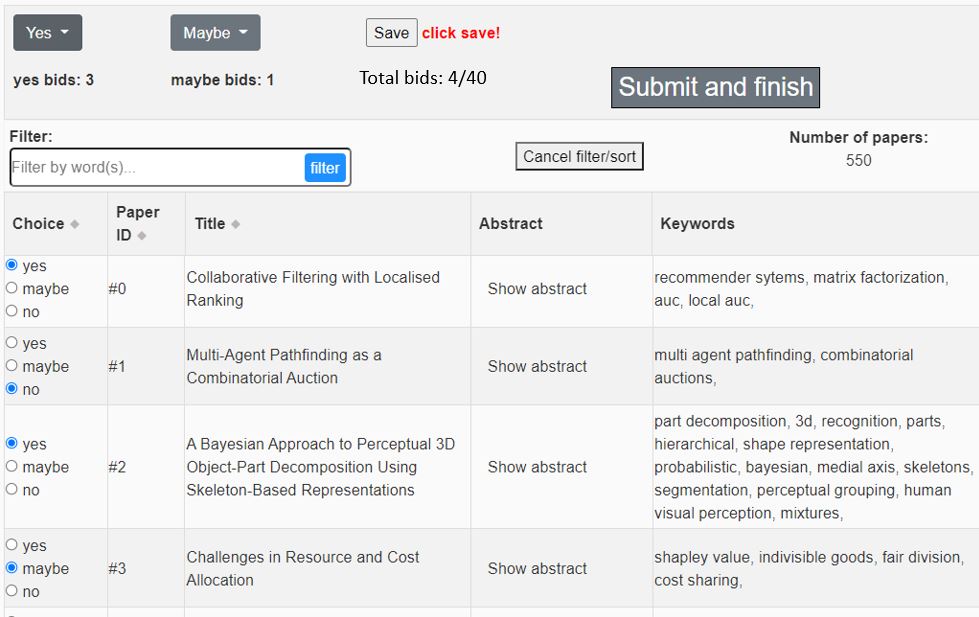}~ ~ ~ ~ ~ ~~~~~~~~~~~~~~~~~~~   \includegraphics[width=0.48\textwidth]{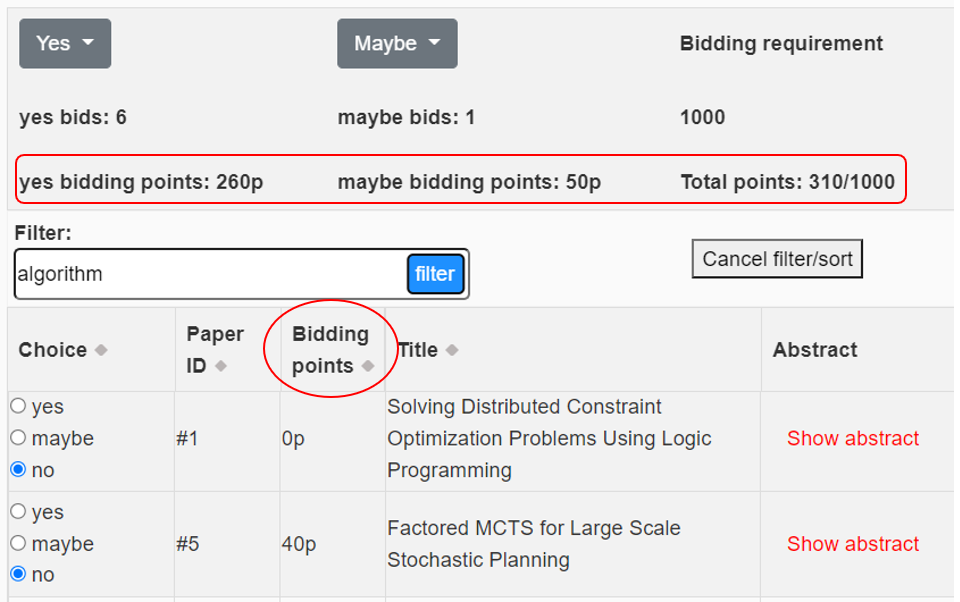}
    \caption{Left: Example of our bidding interface used for all experiments. Right: The interface with the additional column for iPrices (called `bidding points') and the budget.\vspace{-2mm}}
    \label{fig:SC_noframe}
\end{figure*}

We implemented a platform that resembles common paper bidding platforms---mainly EasyChair and ConfMaster.\footnote{See \url{https://easychair.org/} and \url{https://confmaster.net/}.} An example of the interface is shown in Figure \ref{fig:SC_noframe}.

\subsection{The Basic Platform}
In all experiments, the participant is presented with a table containing all papers. For each paper, the table specifies the title and keywords, and the user may click a paper to expand and read the abstract. The user can bid on each paper using a radio button whose states are No/Maybe/Yes, where No is the default option. As is common in bidding platforms, we implemented basic search and filtering capabilities. The user may type a string in order to see only the papers containing this exact string anywhere in the title, keywords, or abstract. At the top, the user also sees how many papers have been marked as Yes and as Maybe so far, and may alter their selection of the papers at any time. Subjects could sort papers according to any column and the initial order depends on the experiment condition.

In some conditions additional information or options were provided in the interface including the inverse price  of papers (called `iPrice' in \citet{meir2021market} and `bidding points' on the platform) or the total bidding requirement, marked in Fig.~\ref{fig:SC_noframe}(Right). We discuss each of these design modifications in their respective section. Following \citet{ShapleyShubik77,meir2021market}, we define the \emph{inverse Price} (iPrice) of a paper $j$ as $p_j:= 100\cdot\min\{1,\frac{r}{d_j}\}$, where $d_j$ is the current number of bids on the paper, and  $r$ is the number of copies of the paper that need to be assigned (throughout this paper $r=3$). Thus a high iPrice indicates current low demand.

\subsection{Field Experiment}\label{sec:FE}
For our field experiment, we used the bidding phase of the COMSOC-2021 international workshop.\footnote{\url{https://comsoc2021.net.technion.ac.il/}} 
We partitioned the set of 42 reviewers randomly into a \textbf{field treatment (FT) group} consisting of 28 people that saw papers' \emph{iPrices} during bidding, and a smaller \textbf{field control (FC) group} of 14 people that saw \emph{no iPrices}.  There were 93 submissions in total.

\paragraph{Bidding Process} Both groups used our platform for bidding, where all 93 submissions were available along with the search and bidding interface shown in Figure~\ref{fig:SC_noframe}. Using this interface, reviewers could also use the platform to report a conflict of interest on papers, but this was scarcely used. The control group had no extra information on demand and were asked to bid positively on at least 12 papers, of which 5-7 will be assigned as in Fig.~\ref{fig:SC_noframe}(Left). The treatment group saw the iPrices as in Fig.~\ref{fig:SC_noframe}(Right), and had a budget of 800 bidding points. These bidding minimums for both groups were purely instructive and were not actively enforced in any way: reviewers could bid on any number of papers. The iPrices were set as explained above and updated on every new login, hence  iPrices were static during a session but may change \emph{between} sessions if an individual reviewer logged back in. We implemented the two caveats recommended in \cite{meir2021market}: (a) the current bidder is always counted as a positive bid on all papers, to prevent price change during the bid; and (b) demands were initialized as uniform rather than empty to prevent a \emph{cold start}. In practice only three reviewers logged in more than once to update their bids. Papers in both groups were initially presented according to their order of submission.\footnote{In hindsight it would have been better to present them in random order, as in the controlled experiments.}

\paragraph{Assignment} The workshop used the bids entered by the committee members as input to a standard utilitarian maximal assignment algorithm with demands and conflict of interests \cite{LianMNW18,garg2010assigning}. The implementation was the same as that of \citet{LianMNW18} which uses Gurobi to solve the assignment ILP and allows for a range of paper and reviewer capacities, each reviewer was assigned 6 or 7 papers. The utility of the overall assignment used, using the bids as a proxy for reviewer utility, was 520.0.  While there may be multiple assignments with the same utility we took the first one that Gurobi provided.

    


\subsection{Controlled Experiments}
In the controlled experiment we had a \textbf{Base (B) group} (same interface as the control group in the field experiment), and several different treatment groups. The main treatments we used were: revealing papers' iPrices to subjects in the \textbf{Price (P) group}; and visually highlighting low-demand papers in the \textbf{Highlight (H) group}. Additional conditions designed to study specific questions will  be explained below. All treatments are between subjects. All subjects faced the same set of 550 papers from AAAI'15, which are publicly available.\footnote{\url{http://www.aaai.org/Library/AAAI/aaai15contents.php}.} Subjects in  group \textbf{B} were requested to bid  on 30 or on 40 papers, of which 8  will be assigned. 


\paragraph{Setting Paper Demand}
As subjects are participating independently of one another, we needed to generate the demand (i.e. the iPrices) for each paper. Rather than generating artificial demand,  we sampled the iPrice directly from a uniform distribution on $[-25,120]$, and truncated to the range $[0,100]$. This is to guarantee we cover the entire range and also have a substantial number of papers with extreme iPrices. Although in reality no paper could have an iPrice of 0 (as it indicates infinite demand), we still wanted to see how this will affect behavior.

\paragraph{Assignment}While the assignment in the controlled experiment plays no role in our analysis, we describe it in Appendix~\ref{apx:assignment} for completeness. Participants were not aware of the exact allocation algorithm, but were told that papers with positive bids were more likely to be assigned, and that the chance also depends on the demand for the paper (to which they may or may not be exposed according to the condition they are in).
The final assignment was displayed to the participant immediately after they submitted their bid, together with the breakdown of the reward.

\paragraph{Incentives}
In  our controlled experiment, participants were not actually reviewing any paper and thus a-priori had no incentive to prefer one paper over another. To mimic the situation of a reviewer trying to select `relevant' papers, we assigned to each participant a set of six `personal keywords' that supposedly reflect her interests. Subjects earned `coins' for each of the 8 papers that were eventually assigned, and how many of these personal keywords they contained (either in the title or in the paper keywords or in the abstract). Each coin increased the bonus  by \$0.25, thereby creating an incentive to bid on relevant papers as common in MTurk Experiments \cite{mason2012conducting}. An important remark is that in real conferences reviewers' interests are often positively correlated. Using common keywords leads to a similar situation in our controlled experiment with a correlation of $0.7\pm 0.16$ in paper relevance among participants.
\rmr{not here:
For $s\in S_i$, we denote by $f^R(s)\in[0,1]$ the normalized reward that $i$ would get if $s$ is assigned (1 being the reward for the most relevant paper). 
}
The personal keywords were selected at random for each participant from the pool of all papers' keywords, with constraints to make sure all participants had a similar amount of relevant papers. \rmr{details in appendix?}
These personal keywords were displayed in a separate box on the screen. 

\paragraph{Instructions and Demo}
To make sure that the (rather complex) instructions of our experiment are understood we: detailed instructions; an online quiz (see appendix); and a demo game. We also informed participants up front that failure to reach minimal required reward may result in rejection of the job---standard for conducting online behavioral research
 \cite{mason2012conducting}. The instructions and quiz focused on explaining that the payment depends only on the assigned papers (8 in total) and not directly on the bid. The demo was very similar to the game except it only contained 50 papers and 3 personal keywords. Participants that did not reach the minimal required reward in the demo could not continue to the game but could try the demo up to 3 times. 
All participants expressed informed consent. 
The study was approved by the IRB of the authors' institution.

\subsection{Measuring Behavior}\label{sec:measures}
Since bidding behavior can be complex and depends on many variables, we develop simple measures that we can compare across subjects and groups of subjects.

For a set of presented papers $S$, we denote by $C(S)\subseteq S$ the subset of papers that were selected by subjects. 
Note that each paper is presented to multiple subject, and counted as a separate `presented paper' for each subject. 
Also 
Note that we treat any positive bid (`Maybe'/`Yes') as a selection. In particular, $C_i$ is the set of papers selected by subject~$i$.

We denote by $\bar C(S):= S\setminus C(S)$ the set of papers from $S$ that were not selected, similarly, $\bar C_i$ are the papers not selected by subject $i$.

We denote by $p_s\in[0,100]$ the iPrice of paper~$s$. In the field experiment, the iPrice was derived from the actual demand as explained above, and was updated with every new login; whereas in the controlled experiment it was generated once per subject and remained fixed.

\begin{figure}
    \centering
 \includegraphics[width=0.46\textwidth]{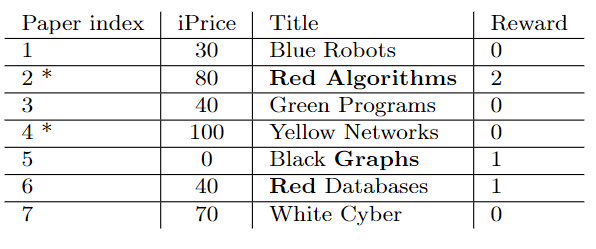}    \vspace{-2mm}
 \caption{Example of calculating reward and sensitivity parameters. Selected papers are marked with *. The reward is  for a subject with the personal keywords  \{\textbf{Red}, \textbf{Graphs}, \textbf{Algorithms}\}.\vspace{-1mm}}
    \label{fig:example}
\end{figure}

\begin{figure}[]
\centering
    
    \includegraphics[width=0.215\textwidth]{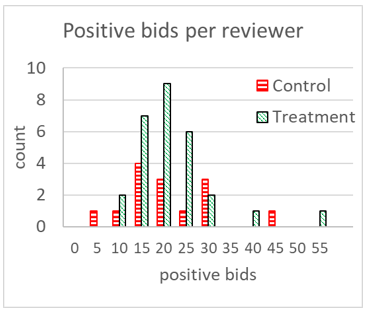}~
           \includegraphics[width=0.262\textwidth]{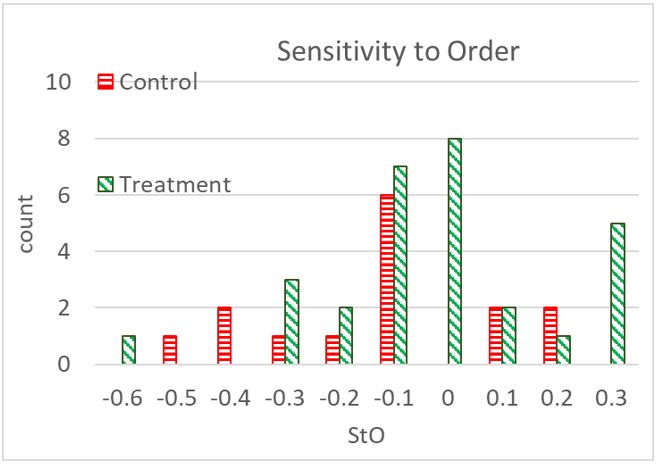}
    \caption{ Left: Histogram of the number of bids for each reviewer in the field experiment. Right: Distribution of individual order sensitivity values in the field experiment.\vspace{-2mm}}
    \label{fig:field_StO}
    \label{fig:field_positive}
\end{figure}
\paragraph{Measuring Individual Behavior}
For each of the features we used $X\in \{R,O,D\}$ (for (R)eward or Relevance, (O)rder,  and (D)emand, respectively) and each paper $s\in S_i$, we denote by $f^X(s)\in[0,1]$ the relevant feature of the displayed paper.

In the example in Fig.~\ref{fig:example} paper \#3 has $f^O(s)=\frac{3}{7}$, $f^D(s)=\frac{80}{100}$, and $f^R(s)=\frac{0}{2}$, as the maximal reward in this example is 2.\footnote{the reward scheme we actually used was a bit different, see instructions in appendix. In particular, the reward for papers with 0 personal keywords, which are most papers, was negative, so there is a strong incentive to avoid them.}

For a subset of samples $S'$, we used the average: $f^X(S'):=\frac{1}{|S'|}\sum_{s\in S'}f^X(s)$.
E.g. for $S'=\{1,2,3\}$ in our example, we have $f^D(S')=\frac{1}{3}(0.3+0+0.8)\cong 0.366$.

For every subject $i\in N$ and feature $X\in\{R,O,D\}$, we defined the `sensitivity-to-X' as the difference between the average value of the feature in selected and unselected papers. Formally:
\labeq{StX}{StX_i:=f^X(C_i) - f^X(\bar C_i).}
$StX_i$ is always in $[-1,1]$, and its expected value is 0 if $i$ is completely insensitive to feature $X$ (e.g. selects papers at random).
For the subject in our example, where the selected papers are $C_i=\{2,4\}$ and $\bar C_i=\{1,3,5,6,7\}$, we have \rmr{maybe move the elaborated example to the appendix?}
\begin{itemize}
    \item $StR = 0.5 - 0.1 = 0.4$, indicating a moderate sensitivity;
    \item $StO = \frac{6}{14} - \frac{22}{35} = -0.2$, meaning the subject tends to select earlier papers; and
    \item $StD = 0.9 - 0.36 = 0.54$, meaning sensitivity towards paper with low demand (=high iPrice).
\end{itemize}

Note that StR cannot be evaluated in the field experiment since we have no direct access to the reviewers' real preferences and expertise. 

\rmr{Is this clear enough and not too mathematical? CSCW can kill us for that} \nm{it is a little too mathmatical -- maybe extend the example a bit more to make it clear? that would tone it down slightly.}
\rmr{and now?}

\paragraph{Measuring Group Behavior}
One way to measure the group behavior is considering the average StX values of group $S$ members (denoted $StX(S)$). When we want to condition on other attributes, we measure the probability of selecting a paper as a function of the relevant feature (e.g. initial position in the table), while controlling for relevance. Formally, given a set of samples $S'$ (say, `all papers in the second quantile of positions that are  highly relevant to their respective subject'), the probability of selection is $PS(S'):=\frac{|C(S')|}{|S'|}$. We can then test if the behavior in two conditions $S,S'$ is different by comparing $StX(S)$ to $StX(S')$ or $PS(S)$ to $PS(S')$, checking if the different is significant using an unpaired t-test.

\section{Results from the Field Experiment}

\rmr{we probably want to start the results on p.6 at most.}
    

\begin{figure}
    \centering
    \includegraphics[width=0.46\linewidth]{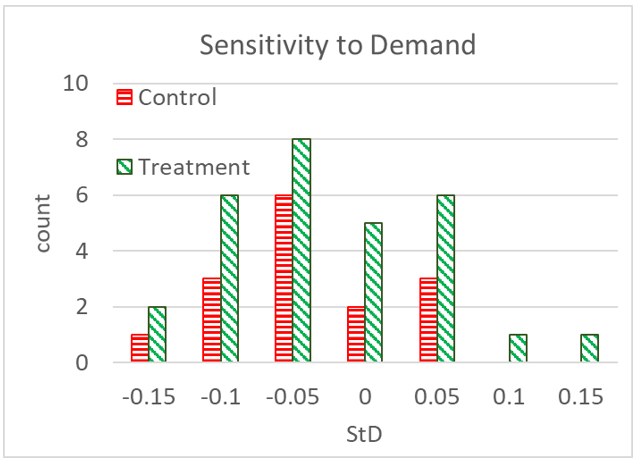}~
    \includegraphics[width=0.46\linewidth]{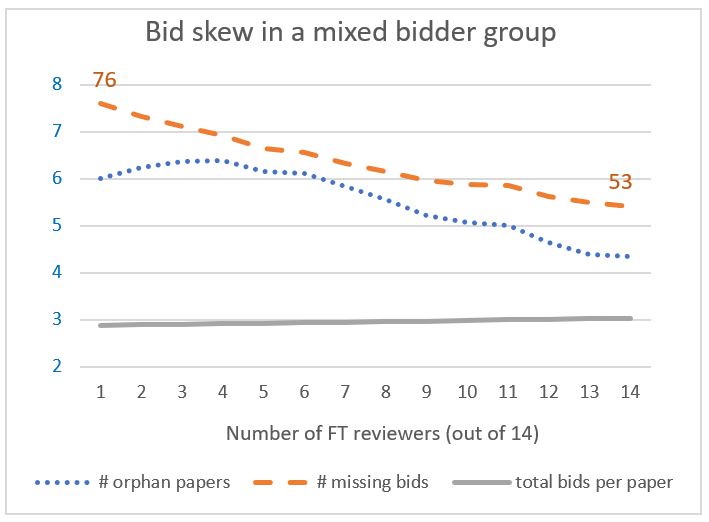}
    \caption{Left: Distribution of demand sensitivity values in the field experiment. Right: Bootstrap results for number of underdemanded and orphan papers in mixed group of 14 reviewers. \vspace{-2mm}}
    \label{fig:field_StD}
    \label{fig:bootstrap}
\end{figure}
\subsection{Distribution of Bids}
The empirical distribution of bids is shown in Figure~\ref{fig:field_positive} (Left). 
In the control group there were a total of 267 bids, 19.1 bids per user, 
 while for the treatment group there were 547 bids, which are 19.5 bids per user. 

 To see if the induced bids in both conditions are drawn from different distributions, we used a two sample Kolmogorov-Smirnov test with the null hypothesis that the treatment distribution was less than the control distribution \cite{hodges1958significance}. This resulted in a test statistic of 0.001429 and a $p$-value of 0.67, so we cannot reject the null hypothesis that average bid amounts are the same.
 
 However what we really want to know is whether bidders where affected by the other factors, in particular order and demand. 
 




 \rmr{Also worth noting that while mean is about the same, the bid amounts are much more concentrated. Need to also see if this holds when bids are weighted by price at time of bidding (indicating more compliance).}

\paragraph{Sensitivity to Order}
According to Table~\ref{tab:counts}, only the control group (FC) demonstrated sensitivity to order, however the effect is barely statistically significant (presumably due to the small number of reviewers on that group). 

\paragraph{Sensitivity to Demand} The first observation from Table~\ref{tab:counts} regarding demand sensitivity is that it is negative in both groups, on average. This may seem surprising but actually makes intuitive sense as in a real conference there is positive correlation in bids, i.e., you are more likely to bid on a popular paper. Hence, only the \emph{difference} between the groups matters.

The average of StD is slightly higher in the treatment group, but this is not statistically significant.  
It is more instructive to look at the distribution of StD values (Fig.~\ref{fig:field_StD}(Left)): we can see clearly that in the treatment group there are several subjects that are highly sensitive to high iPrices (i.e. to low demand), whereas the distribution of the others is similar to the control group.


\subsection{Skewed Bids} 
We compared  the number of papers that were under-demanded in each group, that is, received fewer than the 3 bids necessary to find a good assignment. In the control group there were 47 papers that received fewer than three bids, with 6 of these being papers that received no bids at all. For the treatment group there were only 6 papers that were under-demanded and only a single paper that received no bids. However, this must be partially due to the difference in the size of the two groups. 

To address this we looked both at the number of orphan papers and at the number of \emph{missing bids} (minimal additional bids required so that every paper has at least 3 bids) that would appear under a bootstrap sampling paradigm \cite{bruce2020practical}. To do this we took the set of bids and sampled a ``small committee" from each group  with 14 reviewers in it 1000 times. 
 As we can see in Fig.~\ref{fig:bootstrap}(Right), although the average number of bids remains unchanged, the number of missing bids and orphans drops significantly as we replace FC bidders with FT bidders, indicating that even the small number of demand-sensitive bidders have a substantial effect on the bid skew.

\subsection{Discussion of the Field Experiment}
The initial results from our field experiment suggest that: (1) there seems to be a weak order effect; (2a) there is some fraction of reviewers that are highly sensitive to the demand when given via bidding points and budgets; (2b) this increased sensitivity to demand reduces the number of missing bids and orphan papers; (3) subjects who had budgets were more compliant, possibly due to differences in the UI. 

However the small number of reviewers makes it difficult to make any strong conclusion. In addition some parameters cannot be controlled (such as inherent demand for papers); or were not controlled in our design (such as paper order or displaying the bidding requirement). We therefore turn to controlled experiments to better understand these effects.


\section{Controlled Experiments}\label{sec:controlled}

\paragraph{Conditions}
Our \textbf{base group (B)} was similar to the control group at the field experiment, except that papers where displayed at a random order, and we added the bidding requirement to the UI in order to rule out this as a potential source of differences between groups. See Fig.~\ref{fig:SC_noframe}.

In addition to the base group, we had the following treatments.

\begin{description}
   \item[iPrices (P)] In this condition  (similarly to the FT group in the field experiment) subjects had an additional column titled 'Bidding points' showing papers' iPrices as integers in the range $[0,100]$. The bidding requirement was set as a 'budget' of 1000 points. 
    \item[Highlight (H)]  
In this condition  we did not show the iPrice, but instead highlighted low-demand papers in green (when iPrice is 100) or yellow (when iPrice in [70,99]). 
     \item[iPrices + Sort (PS)] Similar to Condition~P, except papers were initially sorted by increasing demand (decreasing iPrice). 
     \item[iPrices + Highlight + Sort (PHS)] Similar to PS, with also highlighting low-demand papers as in Condition~H.
\item[Implicit Request (IR)] This condition was identical to the base condition, except that the bidding requirement did not appear on the screen during bidding.
\end{description}

\paragraph{Data Collection}
We collected data from 338 participants on Amazon Mechanical Turk.
\rmr{add specific details on AMT demographics} Subjects were allowed to play up to three times. Participants were randomly assigned to the base group or to one of the treatment groups. The total number of participants of each group appears in the second column in Table~\ref{tab:counts}.
The threshold for rejection was set at 12 coins.

\def\hhline{\cline{1-5}\cline{7-9}}

\begin{table*}
\centering
\begin{tabular}{|c|c|c|c|c|c|c|c|c|}

   


\hhline
Code &
Condition & 
participants &
non-spammers& games
 & & StReward  & StOrder  &  StDemand \\
\hhline
B&Base&50&29&29&&$0.34 \pm 0.07$&$-0.11 \pm 0.10$&$-0.01 \pm 0.03$\#\\								
P&iPrices&124&80&80&&$0.34 \pm 0.04$&$-0.16 \pm 0.07$&$0.08 \pm 0.04$\\								
H&Highlight&43&21&39&&$0.36 \pm 0.10$&$-0.13 \pm 0.07$&$0.05 \pm 0.04$\\								\hhline
PS&P+ Sort&34&17&17&&$0.29 \pm 0.06$&$0.02 \pm 0.07$&$0.14 \pm 0.10$\\								
PHS&P+H+Sort&33&28&59&&$0.44 \pm 0.11$&$0.07 \pm 0.06$&$0.09 \pm 0.10$\#\\								\hhline
IR&Imp. Req.&54&36&36&&$0.36 \pm 0.06$&$-0.12 \pm 0.09$&$-0.01 \pm 0.03$\\		  \hhline\multicolumn{9}{}{}\\[-1.7ex] 				\cline{1-5}		
 \multicolumn{2}{|c|}{Total (controlled exp.)}&338&211&260&\multicolumn{4}{c}{}\\	  \cline{1-5}		  \multicolumn{9}{}{}\\[-1.7ex] 				\hhline	
FC&Control&14&14&--&&--&$-0.12 \pm 0.11$&$-0.04 \pm 0.03$\\								
FT&Treatment&28&28&--&&--&$-0.03 \pm 0.08$&$-0.03 \pm 0.03$\\

\hhline
\end{tabular}
\caption{\label{tab:counts} The left side shows number of participants and played games in each group in the controlled experiment. 
The right columns show the average sensitivity of each group (non-spammers only) to each parameter, within 2 standard errors. We mark with \# results in the controlled experiment that do not statistically differ from 0.\vspace{-2mm}}
\end{table*}

\paragraph{Spammers and Sensitivity to Relevance}
There was a distinctive  group of subjects who did not respond to paper relevance (`spammers') and were not included in the rest of the analysis. We explain this in detail in Appendix~\ref{apx:spam}.

\medskip
To better understand the isolated effect of each factor, we start by analysing the Base condition and conditions (H)ighlight and i(P)rices. For Example, the StR column in Table~\ref{tab:counts} shows that in all groups the mean sensitivity (of non-spammers) is about 0.2, and is significantly higher than 0.

\subsection{Paper Order}

\begin{figure}
    \centering
    \includegraphics[width=0.48\textwidth]{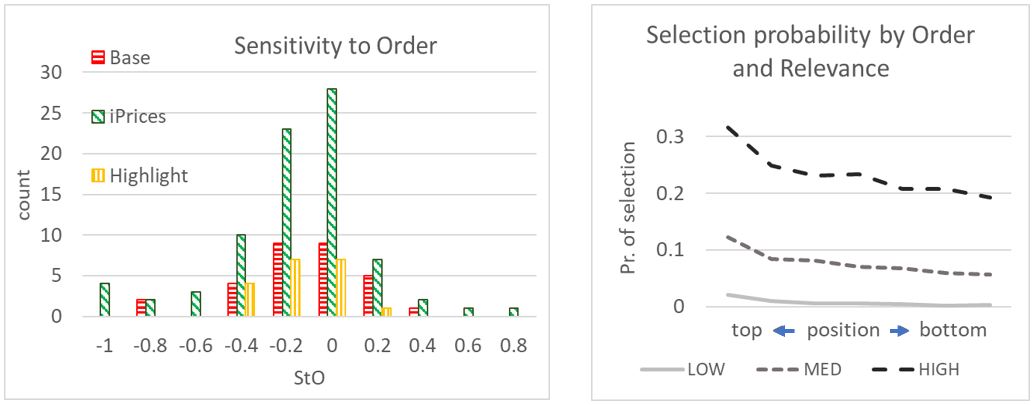}
    \includegraphics[width=.48\textwidth]{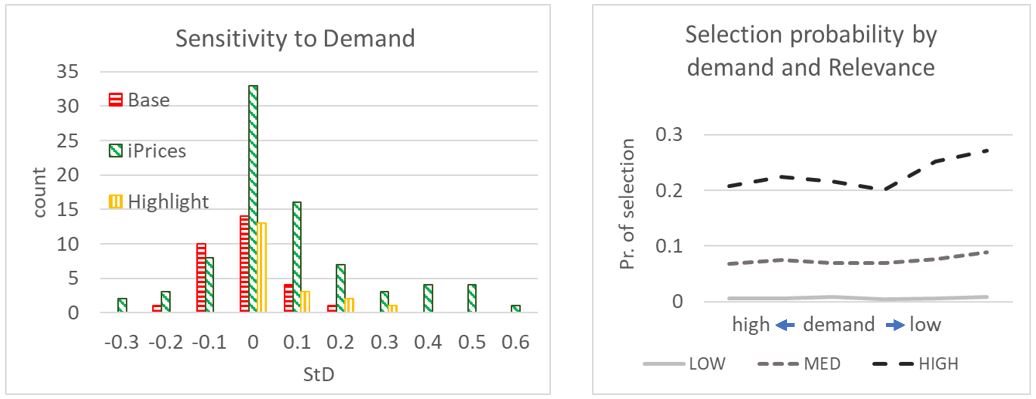}
    \vspace{-0mm}
    \caption{Left: Histogram of individual StO (top) and StD (bottom) values. Right: Selection probability of a paper, conditional on its position and relevance (top), and on its demand and relevance (bottom). Probability is calculated over all subjects in conditions B,P,H. LOW/MED/HIGH relevance means that the paper contained 0, 1, or more relevant words, respectively.\vspace{-2mm}}
    \label{fig:main}
\end{figure}

We can see that in all three conditions, there is  similar average sensitivity to order, of about -0.13, i.e. there is a statistically significant bias to papers that appear earlier. However reward still plays a more important role in selection.\footnote{For many spammer subjects, the StO was even more negative, which is not surprising or interesting.}
But are all subjects slightly biased or is it a small number of highly biased subjects?  For this, we look at the distribution of individual StO values in our controlled experiment (Fig.~\ref{fig:main}, top left).

From the figure, it seems that most subjects are prone to some bias (sensitivity is most often negative but not below $-0.4$); yet there is a non-negligible number of subjects with a very strong sensitivity, which essentially marked papers at the very top. Some subjects had high positive StO values, meaning they deliberately marked papers at the bottom of the list. 

Another question we can ask is whether all papers are equally likely to be promoted when appearing earlier. 
As we can see in Fig.~\ref{fig:main} (top right), primacy affects irrelevant and relevant papers alike, where selection probability drops sharply for papers that are not at the top, and then continues to decrease moderately.


Our findings regarding order effect  are largely consistent with those of \citet{cabanac2013capitalizing} from real conferences, and thus support our Hypothesis~1. The added value of our controlled experiments is two-fold:  how order effects are distributed across the population;    the dependence of order sensitivity (or lack of thereof) on the relevance of the paper. 

\paragraph{Consistency}
A third question we may ask is whether the bias towards early papers is \emph{consistent}. We analyzed the behavior of subjects who played two or three times (all from Condition~H), comparing their StO measure each time.

The between-subject variance of StO is $0.123$---slightly higher than the average within-subject variance of $0.095$. This indicates that participants maintain some  consistency in their sensitivity to order.


\subsection{Paper Demand}
We considered two ways to communicate papers' demand to subjects. The first was adopting the market scheme of \citet{meir2021market} where low-demand papers have high iPrices (condition P). In condition H we simply highlighted the low-demand papers visually. 


\paragraph{Sensitivity to Demand}
The right column in Table~\ref{tab:counts} shows that the Base group is completely insensitive to the demand (as expected, since they have no information about it); the iPrice scheme is moderately effective; and highlighting alone has a small effect (barely statistically significant). 
Looking at the distribution of sensitivity to demand  in Fig.~\ref{fig:main} (bottom left), we can see that in contrast to the primacy effect, most subjects in conditions P and H are not sensitive to the demand.
The effect we see is due to a  relatively small number of highly sensitive subjects. This corroborates our initial finding from the field experiment, and supports our Hypothesis~2. 

\paragraph{Price Scheme More Effective than Highlighting}
We can see in Table~\ref{tab:counts} that the effect of highlighting papers by itself is borderline significant (only 3 of the 21 subjects demonstrated significant bias towards highlighted papers in Fig.~\ref{fig:main}). In contrast, about third of the subjects who were exposed to iPrices were significantly affected, and the overall bias doubled. 




\paragraph{Which Papers are Affected?}
We can see that the effect of high iPrices is mainly on papers that are already relevant (Fig.~\ref{fig:main}, bottom right). This is another difference from the effect of paper order. It is also another evidence of rational decision making (in the economic sense), as the iPrice indicates the probability of getting the paper.

\rmr{\cite{stoddard2015popularity} talks about popularity vs. quality. perhaps relevant to claim that just promoting arbitrary low-demand papers is less beneficial since they may be less relevant to the bidders.}
\rmr{check if all these bids come from demand insensitive subjects}




\paragraph{Consistency}
Similarly to order effects, subjects who played 2 or 3 games exhibit some consistency in their sensitivity to demand, with a between-subject variance of $0.022$ vs. $0.015$ within-subject.

\rmr{need to do this with P group!}


\subsection{Using All Treatments?}
Since paper order, iPrice and highlighting all have some positive effect, it might make sense to combine them together in order to influence people to spread their bids even more.
In particular, we can adopt the suggestion of \citet{cabanac2013capitalizing} to actively sort papers by increasing demand \emph{and} combine this with \citet{meir2021market} iPrice scheme. 
Further, we can highlight the most underdemanded papers in the table.
We therefore ran another experiment with two more groups: In group P+Sort we displayed iPrices and budget as in condition P \emph{and} sorted the papers initially by decreasing iPrice (so underdemanded papers are on top); In group P+H+Sort we did the same \emph{and} highlighted the underdemanded (high-iPrice) papers as in condition H.  Note that in these two conditions StO is the same as (negative) StD. 


By looking back at the StD results in Table~\ref{tab:counts},  we can see that in group P+Sort there is more sensitivity to demand, as expected. It is also higher than the mean StO of subjects in the previous conditions, indicating some synergy between the price scheme and the initial order.
However highlighting papers in addition did not help at all (see condition P+H+Sort), and in fact only added noise. We analyse possible reasons for this phenomenon in Appendix~\ref{apx:PHS}, showing it might be an artifact of a single (or few) AMT participant connecting from multiple accounts.

\rmr{We already embedded the interpersonal differences analysis in the text}





\subsection{Compliance with Bidding Instructions}
A very visible difference between the groups at the field experiment (Fig.~\ref{fig:SC_noframe}(Left)) is that while the average amount of bids per reviewer is similar, the treatment group seems more concentrated around the mean. 

 To more accurately measure this difference we define a new measure called \emph{Compliance Ratio}. The full details are in Appendix~\ref{apx:compliance} but intuitively a compliance ratio of 1 means exactly exhaust the budget (in conditions with iPrices), or bidding exactly on the required number of papers (when there are no iPrices).

A number below 1 indicates underbidding and a greater number indicates overbidding. 

Indeed the compliance ratio  was concentrated around 1 only in condition FT and not FC, but we suspected that the difference was only because in condition FT the budget was displayed throughout (in FC the requirement only appeared in the instructions, as in most conferences).

Our controlled experiment confirmed this hypothesis:
the IR condition is identical to the Base condition, except for the fact that the bidding requirement was not displayed (as in FC). Indeed, the result was that while in both cases the \emph{average compliance} was close to 1, values were much more concentrated in Condition~B, whereas many participants over- and under-bid in Condition~IR.

\section{Discussion}
Our combined experiments in bidding behavior show that:
\begin{enumerate}
    \item Bidding likelihood increases uniformly for papers appearing higher  in the list (corroborating previous empirical findings); 
    \item Presenting papers' demand in the form of iPrices positively influences a small but non-negligible subset of people to shift their selection to low-demand papers;
    \item Presenting the bidding requirement during bidding (rather than just include it in the instructions beforehand) results in much higher compliance. 
\end{enumerate}


Our field experiment further showed that shifting the demand of even few bidders towards low-demand papers, reduces the skew in bids and makes sure more papers get the minimal required amount of bids.

\paragraph{Critique on experimental results}
There are two main concerns about the validity of our results. 
First, there is an internal validity issue: One can ask whether the behavior we see is consistent or sporadic. This is important as consistency also means predictability. Our preliminary analysis shows that subjects exhibit at least some level of consistency but this should be studied more in-depth over longer time periods and with diverse input.

Another concern is external validity: will the behavior of researchers bidding on real papers be similar to that of AMT workers who play a game for recreation and/or money?\footnote{The analysis in Appendix~\ref{apx:PHS} raises one such concern.} 

We argue that the answer is \underline{yes}. While it is  clear that the \emph{preferences} of actual reviewers over real paper assignment are very different from those of AMT participants in our controlled experiment, it is much more likely that both groups demonstrate the same \emph{behavioral biases and tendencies} in trying to obtain their preferred outcome.\footnote{Note that we restricted our AMT participants to similar demographics by requiring a university degree.}

In that respect, our use of AMT is similar to its use in consumer behavior research, where controlled experiments with simulated (rather than actual) purchases are used to complement field studies and deepen understanding~\cite{ghose2014examining}.

More generally, results from AMT experiments are considered reliable despite some differences in personality traits~\cite{goodman2013data}, especially if subjects are filtered based on their comprehension of the task (as we do).


\paragraph{Critique on paper bidding with iPrices}
There are several concerns raised by the suggested bidding scheme in \cite{meir2021market}. Mostly regarding fair treatment of papers and  strategic considerations of bidders (e.g. is it better to bid earlier or later). \citet{meir2021market} directly address most of these concerns in the original paper, where their main point is that bidders are free to ignore instructions and behave as they would without demand information, but any bidder that does take this information into account improves the outcome both for herself and for the others.

We can also add that we did not encounter any adverse effects in our field experiment. However we should keep in mind it was in a small scale.

Another possible objection is that automated matching enabled by systems like TPMS makes bidding redundant altogether, or at least less important. That may be true in the future but as shown in \cite{fiez2020} (see our Introduction), current automated fit-scores are also highly skewed, and may therefore exacerbate the problem rather than solve it.

\paragraph{Practical Recommendations}
We believe that adopting the simple market scheme of \citet{meir2021market} can have a positive influence on distribution of bids during bidding phase. This influence can be increased by combining other UI factors such as highlighting and/or use the current demand as a factor in sorting presented papers~\cite{cabanac2013capitalizing,fiez2020super}. Regardless of the bidding scheme, we recommend that the bidding requirement (in terms of number of positive bids or budget) will be displayed during bidding. These changes can be easily implemented in existing platforms such as EasyChair and ConfMaster, and be offered to conference organizers as optional features. 

We recommend doing these changes carefully:
\begin{itemize}
    \item Consult UX/UI experts regarding the best way to highlight papers so as to avoid confusion, 
    choosing the best terms to describe iPrices and budgets, etc.;
    \item Explain reviewers/committee members that they can bid as they wish (even ignore all additional information), but will be more likely to get their desired papers by following the bidding instructions;
    \item 
As for paper order, we should keep in mind that most platforms offer the user flexibility in how to sort the papers, so users should have to option to choose  whether demand should be a factor in this order;
    \item Test suggested changes on a subset of conference participants and/or in smaller workshops before full adoption.
\end{itemize}
We hope these suggestions will contribute to improving the review process for all. 



\begin{acks}
Nicholas Mattei was supported by NSF Awards IIS-RI-2007955, IIS-III-2107505, and IIS-RI-2134857, as well as an IBM Faculty Award and a Google Research Scholar Award.
\end{acks}

\bibliographystyle{ACM-Reference-Format} 
\bibliography{reviews}
 \appendix
 \onecolumn

\begin{figure}
    \centering
    \vspace{-2mm}
\includegraphics[width=0.40\textwidth]{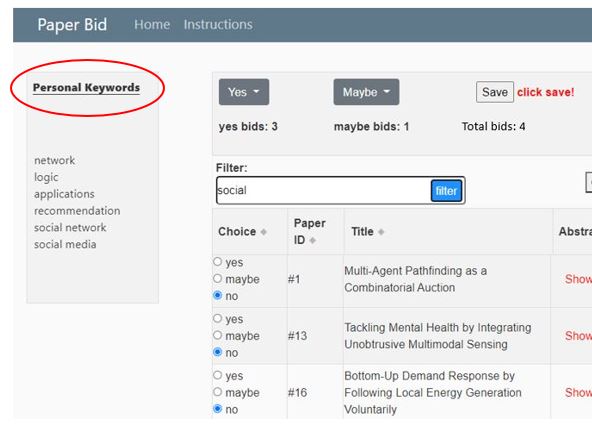}
    \caption{A screenshot of the controlled experiment. The bidding interface  is the same as in the field experiment. The personal keywords representing the subject's interests are highlighted on the left.
     \label{fig:frame}}
\end{figure}
\section{Interface and Assignment Algorithm in the Controlled Experiment}\label{apx:assignment}
The interface is shown in Fig.~\ref{fig:frame}, where we can see the personal keywords on the left.

Recall that in the controlled experiment, there is only one player at a time, but every paper as a fixed iPrice (see Section~\ref{sec:measures}) that reflects its demand.

After selection, we determined the assignment based on subject's bids using a simple randomized allocation: The iPrice of each paper  is used as an approximate assignment probability $p_j$ (see \cite{meir2021market}). 

The algorithm first considers all 'Yes' bids in random order and assigns each paper with probability $p_j$, until $8$ papers are assigned or until all of `Yes' bids have been considered. 

It then repeats the process for `Maybe' bids, and then for all other papers.


\section{Spammers and Sensitivity to relevance}\label{apx:spam}

\rmr{this part is more about experiment design than bidding behavior so perhaps also move to appendix}
The first thing we had to check was whether subjects are responsive to incentives at all, as otherwise there is not much value in measuring their behavior. 

We had an internal criterion used to filter out participants that had consistently failed to select relevant papers.  For this, we calculated for each subject her sensitivity to relevance (StR).

Note that a completely random selection should result with StR that is close to 0, whereas even a very weak tendency to select relevant papers (e.g. avoiding papers that have no personal keywords) should result in a strictly positive StR. 

\begin{figure}
    \centering
    \includegraphics[width=0.75\textwidth]{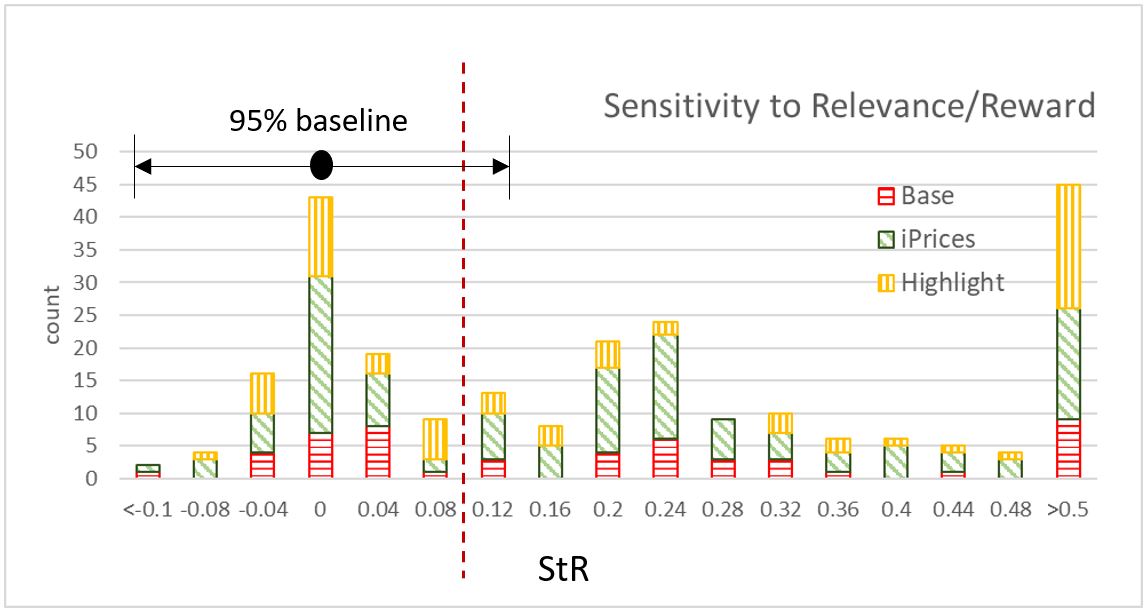}
    \caption{A histogram of subjects' StR. Subjects of different conditions are separated by color and pattern.  The arrow indicates the range in which we expect to find subjects who are completely insensitive. This is the 95\% confidence interval for the StO value of a subject that picks at least 20 papers completely at random. Subjects who pick fewer papers have a higher chance of exceeding the interval.
    \label{fig:S2R}}
\end{figure}

The empirical distribution of StR is visible in Fig.~\ref{fig:S2R}. We can see that there is a bimodal distribution, indicating two types of subjects: a large minority of subjects whose StR is distributed around 0 (left of the dashed line in the figure), and the majority of subjects, whose StR is well separated from 0.

We treat the first type as 'spammers', and do not include them in our behavioral analyses.\footnote{A participant may be a `spammer' either if she plays without effort for a quick gain, or if she failed to understand and follow the reward structure. Some evidence suggest that most spammers are of the first type, as they had spent much less time on bidding, and provided unreasonable demographic information.} The right column in Table~\ref{tab:counts} shows the number of usable subjects from each group. Note that our criterion for spammers is conservative, and that there may still be some spammers within the low-StR subjects we analyze. 

\paragraph{Payment Rejection}We used a more conservative criterion do reject AMT subjects. 

 We only rejected subjects who failed the announced  criterion (at least 12 coins) \emph{and} our internal `spammer' criterion explained above. Thus spammers who (by chance or due to false classification) reached 12 coins were paid, but their data was not included in the analysis.

\section{Strange Behavior in the P+H+Sort Condition}\label{apx:PHS}
Recall that the mean sensitivity to demand in Condition~PHS was substantially lower than in PS alone, even though the conditions are very similar, and Highlighting should only increase the sensitivity. 

\begin{figure}
    \centering
\includegraphics[width=0.42\textwidth]{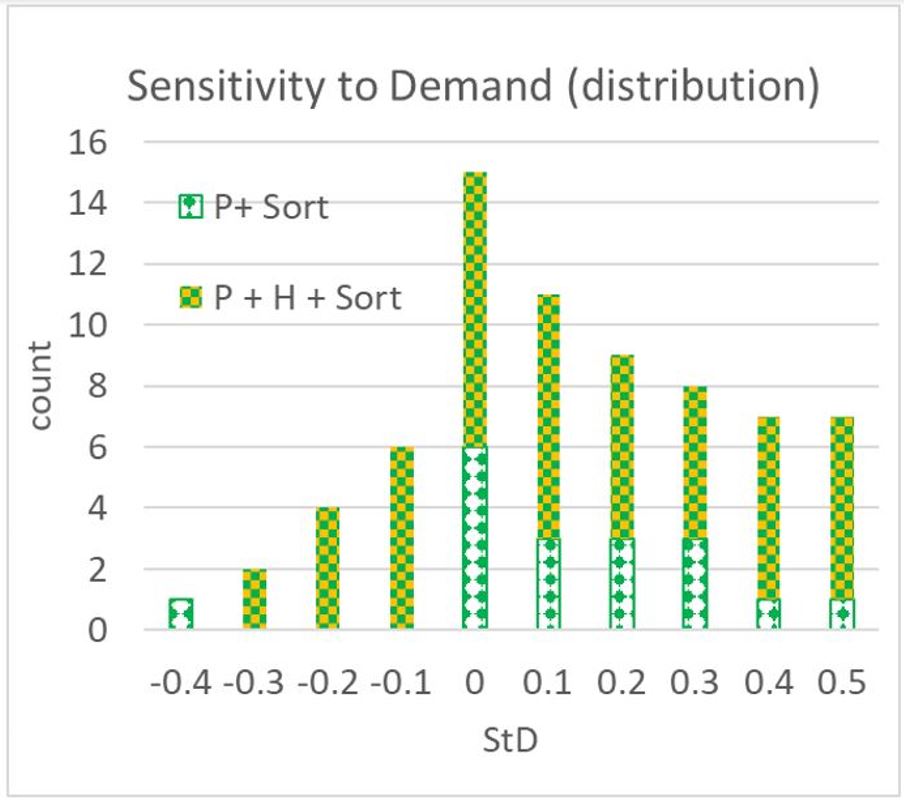}   
\caption{ Distribution of individual StD values in the two new conditions.}
    \label{fig:dist_HPS}
\end{figure}

We can see a possible reason for this in the distribution of personal StD values in Fig.~\ref{fig:comp}(Right): while almost all subject in group P+H+Sort were influenced, it seems that some of them deliberately selected \emph{high-demand} papers! Moreover, for those subjects who played more than one game, this behavior was consistent. 

While it is possible that our instructions in this condition only were not clear and led to confusion, there is also an alternative explanation. 

An `exact bidder' is a subject who bids exactly the budget. While this behavior was rare in general (occurred in about 10\% of the games in the other conditions), in the PHS condition more than half the games were exact. Moreover, there were 11 (out of 28) users who were exact in all of their games. 

While exact bidding is not a problem on its own, such an excessive rate of two rare behaviors (exact bidding and negative StD) in a single condition may indicate that there were few persons (or perhaps even one) behind many of the accounts. 

While we cannot verify this conjecture, it is a reminder about how careful we should be when designing and analyzing online experiments, and  of the need for independent corroborations. 

When ignoring games with suspicious exact bidding, we get mean StD rate of $0.13\pm 0.09$ and  $0.15 \pm 0.09$ in the PS and PHS conditions, respectively. 
This indicates that the PHS condition is indeed the most effective in making bidders demand-sensitive are reduce bid skew.

\section{Compliance with Bidding Instructions}\label{apx:compliance}
A very visible difference between the groups at the field experiment is that in the treatment group the total bid amounts of each reviewer are \emph{concentrated} (with a few outliers), whereas in the control group the distribution is scattered. This is even though the control group were instructed to bid on a given number of papers (12), and the instructions to the treatment group was in terms of bidding points. To more accurately measure this difference we define a new measure.

The \emph{compliance ratio} of a subject with instructions to bid on $R$ papers is computed as $Comp_i:=\frac{|C_i|}{R}$. For subjects that have prices and budgets we compute the compliance as $Comp_i:=\frac{\sum_{s'\in C_i}p_s}{B}$ where $B$ is the budget.

In either case, a number below 1 indicates underbidding and a greater number indicates overbidding. 

Indeed, for condition~FC (control) the average compliance ratio was 1.59 (median 1.55) with a standard deviation of 0.75; whereas in condition FT most subjects were highly compliant, with average, median, and standard deviation of 1.15, 0.97, and 0.52 respectively.  

However we suspected that the difference was not due to the use of bidding points:
Only the subjects in the treatment group can easily track their progress, through the budget that is shown throughout bidding (see Fig.~\ref{fig:SC_noframe}(bottom)). 

We therefore formulated a third hypothesis:
\begin{description}
   \item[Hypothesis~3]: Displaying the bidding requirement (regardless of other information) makes reviewers more compliant.  
\end{description}
Note that by `more compliant' we mean the compliance ratio of more subjects is closer to 1, not necessarily higher. 

\paragraph{Controlled experiment}
To test this hypothesis, we contrasted the Base condition with the `Implicit Request' (IR) condition.  In the IR condition  (as in most conferences and in our FC group) the bidding requirement was not part of the interface, but still appeared on the instructions that were accessible throughout the experiment. 

We can see in Table~\ref{tab:counts} that groups B and IR behaved similarly in terms of our three sensitivity measures. 

\begin{figure}
    \centering
    \includegraphics[width=0.46\textwidth]{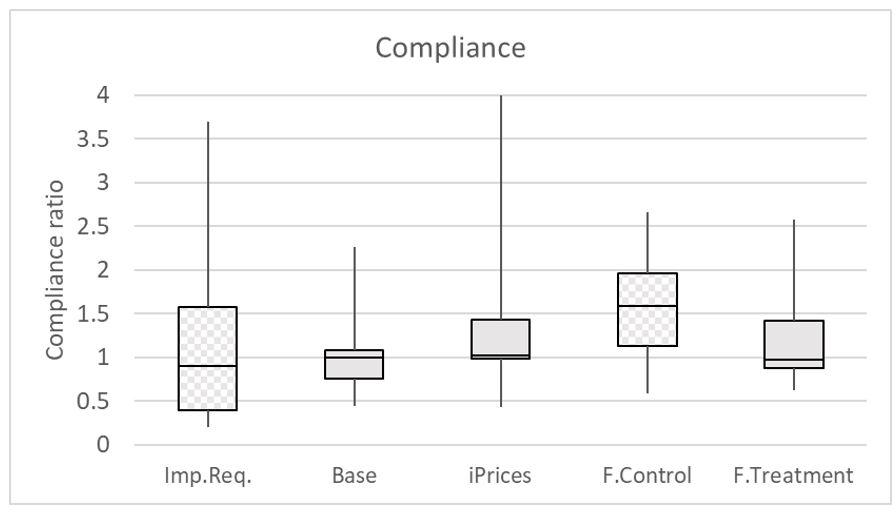}
    \caption{A box plot of the compliance ratio distribution under different conditions (showing quartiles 1-3 and top and bottom 5 percentiles).  In the checkered conditions (IR,FC) the bidding requirement was not shown during bidding.\rmr{wrong graph???}
    }
    \label{fig:comp}
\end{figure}

\paragraph{Results}
Comparing groups IR and B allows us to directly test Hypothesis~3. Indeed, we can see in the two leftmost columns of Fig.~\ref{fig:comp} that merely presenting the bidding requirement on the screen substantially boosts compliance: while the average compliance in both is close to 1 (1.14 and 1.07, respectively), the variance of IR is much higher ($p$-value is 0.001 in an F-test). This finding supports the hypothesis.  Recall that subjects could view the instructions at any time 
, so the strength of the effect is rather surprising.

There was no substantial difference between groups P and B in terms of concentration, although there was less underbidding and more overbidding in group~P. This is in contrast to what we saw at the field experiment, but recall that groups FC was more similar to IR than to B in this respect. 

\section{Experiment Instructions and Quiz}\label{apx:quiz}

The reader is encouraged to try the interface at the following link (this link may not be stable, will provide a stable link at the final version):

\url{https://paperbid.herokuapp.com/registerA/}
\includepdf[pages=-]{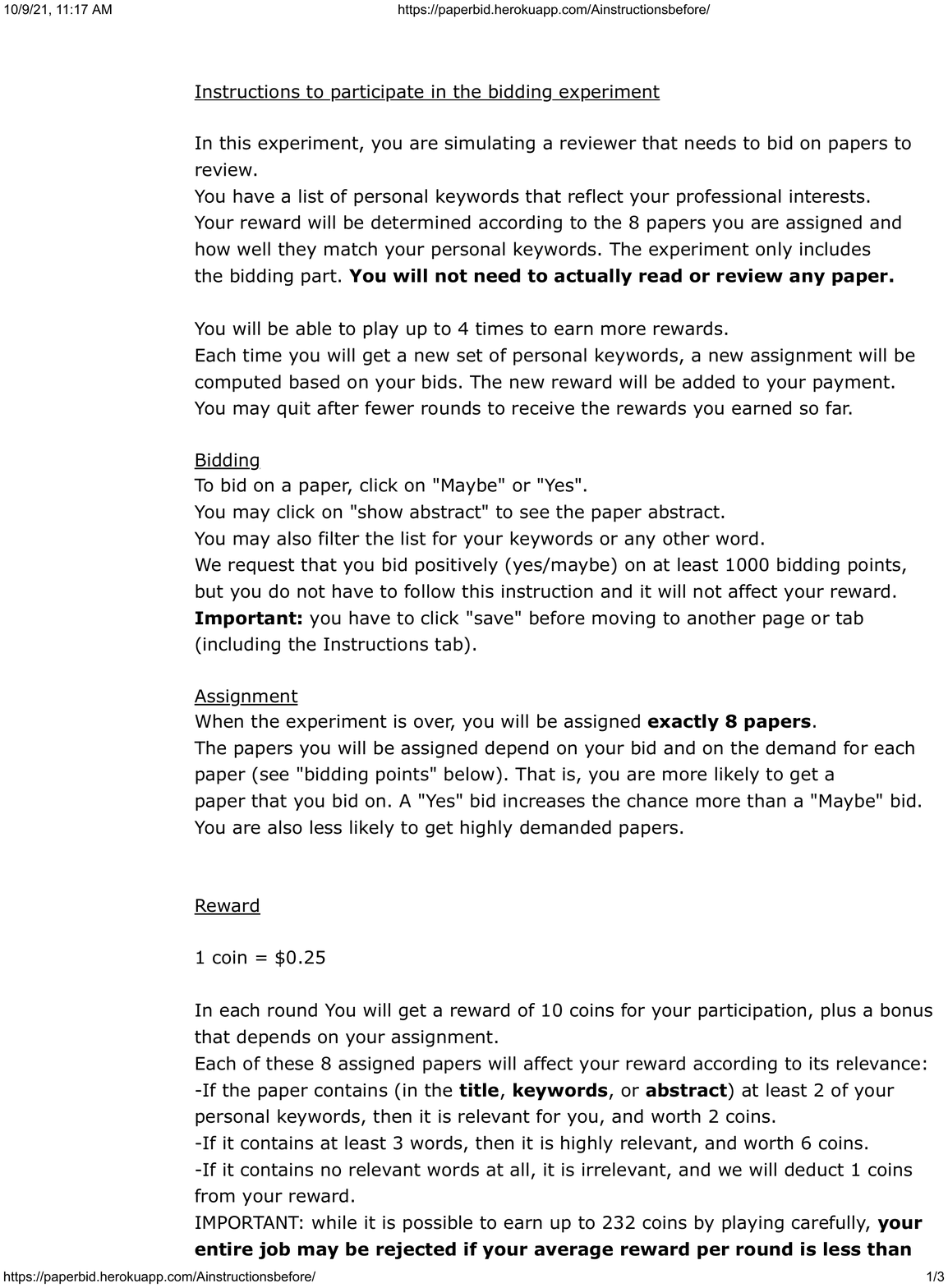}
\begin{figure}
    \centering
    \includegraphics[width=\textwidth]{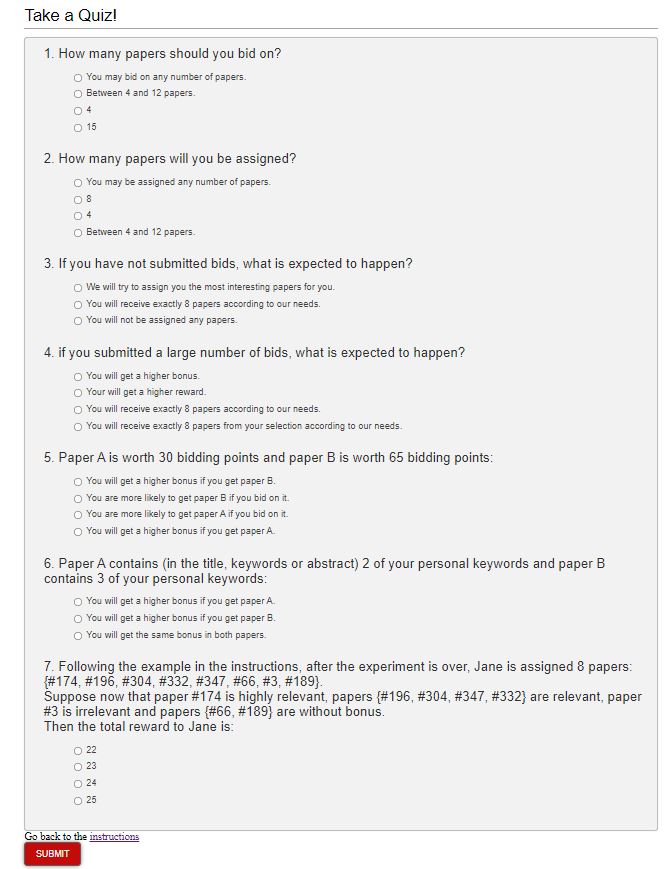}
    \caption{}
    \label{fig:my_label}
\end{figure}
\end{document}